\documentclass{aa}
\usepackage{epsfig}
\usepackage{natbib}
\usepackage{graphics}
\usepackage{color} \usepackage{ulem}
\usepackage{color}
\newfont{\tss}{cmssdc10 scaled 950}

\begin{document}

\title{VLT/X-shooter observations of the low-metallicity blue compact dwarf
galaxy PHL 293B including a luminous blue variable star\thanks{Based on 
observations collected at the European Southern Observatory, Chile, ESO program 
60.A-9442(A).}\thanks{The reduced data in Figures 1 and 2 are available 
at the CDS via anonymous ftp to cdsarc.u-strasbg.fr (130.79.128.5)
or via http://cdsweb.u-strasbg.fr/viz-bin/qcat?J/A+A/533/A25}
}

\author{Y. I.\ Izotov \inst{1,2,3}
\and N. G.\ Guseva \inst{1,2}
\and K. J.\ Fricke \inst{1,4}
\and C.\ Henkel \inst{1}
}
\offprints{Y.I. Izotov, izotov@mao.kiev.ua}
\institute{          Max-Planck-Institut f\"ur Radioastronomie, 
                     Auf dem H\"ugel 
                     69, 53121 Bonn, Germany
\and
                     Main Astronomical Observatory,
                     Ukrainian National Academy of Sciences,
                     Zabolotnoho 27, Kyiv 03680,  Ukraine
\and
                     LUTH, Observatoire de Paris, CNRS, 
                     Universite Paris Diderot,
                     Place Jules Janssen 92190 Meudon, France
\and
                     Institut f\"ur Astrophysik, 
                     G\"ottingen Universit\"at, Friedrich-Hund-Platz 1, 
                     37077 G\"ottingen, Germany
}

\date{Received \hskip 2cm; Accepted}

\abstract
{ We present VLT/X-shooter spectroscopic observations 
in the wavelength range $\lambda$$\lambda$3000 -- 23000\AA\ of the extremely
metal-deficient blue compact dwarf (BCD) galaxy PHL 293B
containing a luminous blue variable (LBV) star and compare them with previous 
data.} 
{This BCD is one of the two lowest-metallicity galaxies where
LBV stars were detected, allowing us to study the LBV 
phenomenon in the extremely low metallicity regime.}  
{We determine abundances of nitrogen, oxygen, neon, 
sulfur, argon, and iron by analyzing the fluxes of narrow components 
of the emission lines using empirical methods and study the properties 
of the LBV from the fluxes and widths of broad emission lines.} 
{We derive an interstellar oxygen abundance of 12+log O/H = 7.71 $\pm$ 0.02,
which is in agreement with previous determinations.
The observed fluxes of narrow Balmer, Paschen and Brackett hydrogen lines 
correspond to the theoretical recombination values after correction for  
extinction with a single value $C$(H$\beta$) = 0.225. This implies that 
the star-forming region observed in the optical range is the only source of 
ionisation and there is no additional source of ionisation that is seen in the 
NIR range but is hidden in the optical range.
We detect three $v$ = 1-0 vibrational lines of molecular hydrogen. 
Their flux ratios and non-detection of $v$ = 2-1 and 3-1 emission lines
suggest that collisional excitation is the main source producing H$_2$ lines.
For the LBV star in PHL 293B we find broad emission with P Cygni profiles 
in several Balmer hydrogen emission lines and for the first time in several 
Paschen hydrogen lines and in several He {{\sc i}} emission lines, 
implying temporal evolution of the LBV on a time scale of 8 years.
The H$\alpha$ luminosity of the LBV star is by one order of magnitude higher
than the one obtained for the LBV star in NGC 2363$\equiv$Mrk 71 which has a 
slightly higher metallicity 12+logO/H = 7.87. The terminal velocity 
of the stellar wind in the low-metallicity LBV of PHL293B is high, 
$\sim$ 800 km s$^{-1}$, 
and is comparable to that seen in spectra of some 
extragalactic LBVs during outbursts. We find that the averaged terminal 
velocities derived from the Paschen and He {{\sc i}} emission lines are by 
some $\sim$ 40 -- 60 km s$^{-1}$ lower than those derived from the Balmer
emission lines. This probably indicates the presence of the wind accelerating 
outward.
}
{}
\keywords{galaxies: fundamental parameters -- galaxies: starburst -- 
galaxies: ISM -- galaxies: abundances -- stars: activity}
\titlerunning{VLT/X-shooter observations of the low-metallicity BCD PHL 293B}
\authorrunning{Y.I.Izotov et al.}
\maketitle


\section{Introduction}

Critical evolutionary phases in the life of massive stars such as the stage
of luminous blue variables \citep[LBVs; ][]{C84} are hard to study due to the
exceedingly short time scales involved. To enhance the chance of
catching the LBV phenomenon, it is best to observe galaxies known to 
contain many massive stars. Blue compact dwarf (BCD) galaxies are suitable  
objects for such studies as their star-forming regions
can harbor up to thousands of massive stars \citep[e.g., ][]{TI05}. 
As the mass loss rate depends on the metallicity, 
BCDs of varying metal deficiency are excellent laboratories to 
test metallicity effects on mass loss \citep{I07}. 

The prevailing belief is that the 
efficiencies of stellar winds in massive stars are significantly reduced at 
low metallicities. However, there is growing
evidence suggesting that massive stars do have stellar winds, 
even in the most metal-deficient BCDs known, with 
interstellar oxygen abundances 
12 + log O/H $\la$ 7.6. \citet{I97} and \citet{L97} discovered
a Wolf-Rayet (WR) stellar population in I Zw 18
with an oxygen abundance 12+log O/H = 7.17 $\pm$ 0.01 \citep{I99}. 
Furthermore, \citet{TI97} have observed stellar P Cygni profiles for the 
Si {\sc iv} $\lambda$1394, $\lambda$1403 absorption lines in the spectrum of 
the emission-line galaxy SBS 0335--052E
with an interstellar oxygen abundance of 
12 + log O/H = 7.30 $\pm$ 0.01 \citep{I07}.

Although we do not know the metallicities of these stars, it appears 
that they were formed in the interstellar medium
with an extremely low metallicity and do possess stellar winds.
The properties of these stellar winds may significantly differ
from those of their high-metallicity counterparts \citep[e.g., ][]{GIT00,CH06}.

To find more objects with observational signatures of massive star 
outflows, \citet{I07} have assembled 
a sample of about 40 emission-line dwarf galaxies which exhibit 
broad components in their strong emission lines, mainly in H$\beta$, 
[O {\sc iii}] $\lambda\lambda$ 4959, 5007, and 
H$\alpha$. Except for four objects which appear to 
contain massive black holes \citep{IT08}, 
the broad emission of all the other objects can be 
attributed to some evolutionary stage of massive stars and to 
their interaction with the circumstellar and interstellar medium, e.g.
WR stars, supernovae, superbubbles or LBV stars.
The attention of \citet{I07} 
was drawn to the spectrum of the BCD PHL 293B $\equiv$ SDSS J2230--0006, 
with 12 + log O/H = 7.72, which shows broad Balmer hydrogen emission lines
with P Cygni profiles, that are characteristic of a  
LBV star. Several months after this finding, \citet{P08} 
discovered another bright LBV in the BCD DDO 68 with 
interstellar 12 + log O/H = 7.14.
The study of these newly discovered LBV stars 
was the focus of the paper by \citet{IT09}.
 
The LBV phase represents a transition in the late 
evolution of all stars with initial masses greater than about 50 $M_\odot$ 
from the stage of O stars burning hydrogen on the 
main-sequence to that of core-helium burning WR stars.
During this phase, the stars lose large amounts 
of mass in recurrent explosive events. In the HR diagram, LBVs lie just to the 
left of the Humphreys-Davidson limit \citep{HD94},
beyond which no stars are observed. They define the locus of an 
instability that prevents further redward evolution, e.g. the 
``strange mode instability'' \citep{K93}.

The exact mechanism giving rise to a LBV outburst is still unknown
\citep[e.g., ][]{MM00,M03}, even though 
many LBVs at solar metallicity have been studied \citep[e.g. ][]{C05}. 
The situation is worse
at low metallicity. \citet{S10} noted that LBVs are often detected in spiral 
galaxies but rarely in dwarf galaxies and the metallicities of most 
of these dwarf galaxies are still unknown. 
Exceptions are the Small Magellanic Cloud with 12+log O/H $\sim$ 8.0 and
the dwarf cometary galaxy NGC 2366 with 12+log O/H = 7.87. In the latter galaxy,
the LBV star has been investigated in detail \citep{D97,D01,P06}. 
Repeating high-quality spectroscopic observations of LBV stars
in more extremely metal-deficient galaxies should help to  
better understand how the LBV phase depends on metallicity.    

We present here new VLT/X-shooter spectroscopic observations of 
PHL 293B over a larger wavelength range 
of $\sim$ $\lambda$3000-23000\AA\ as compared to the SDSS 
($\sim$ $\lambda$3800-9200\AA) and VLT/UVES 
($\sim$ $\lambda$3000-7000\AA)
optical data discussed by \citet{IT09}. 
These new observations allow us
to derive element abundances in the H {\sc ii} region, to study whether any
hidden star formation is present in this galaxy, and to obtain 
emission line parameters of the LBV star for the epoch $\sim$ 8 years 
after previous observations. 
The new observations are described in Sect. \ref{obs}. 
In Sect. \ref{host} we discuss the properties of the host galaxy.
We derive the element abundances, discuss extinction and hidden 
star formation and excitation mechanisms
of the H$_2$ emission in the NIR range. 
The properties of the 
LBV broad emission are analysed in Sect. \ref{lbv}.
Our conclusions are summarised in Sect. \ref{concl}.

\begin{figure*}[t]
\hspace*{0.0cm}\psfig{figure=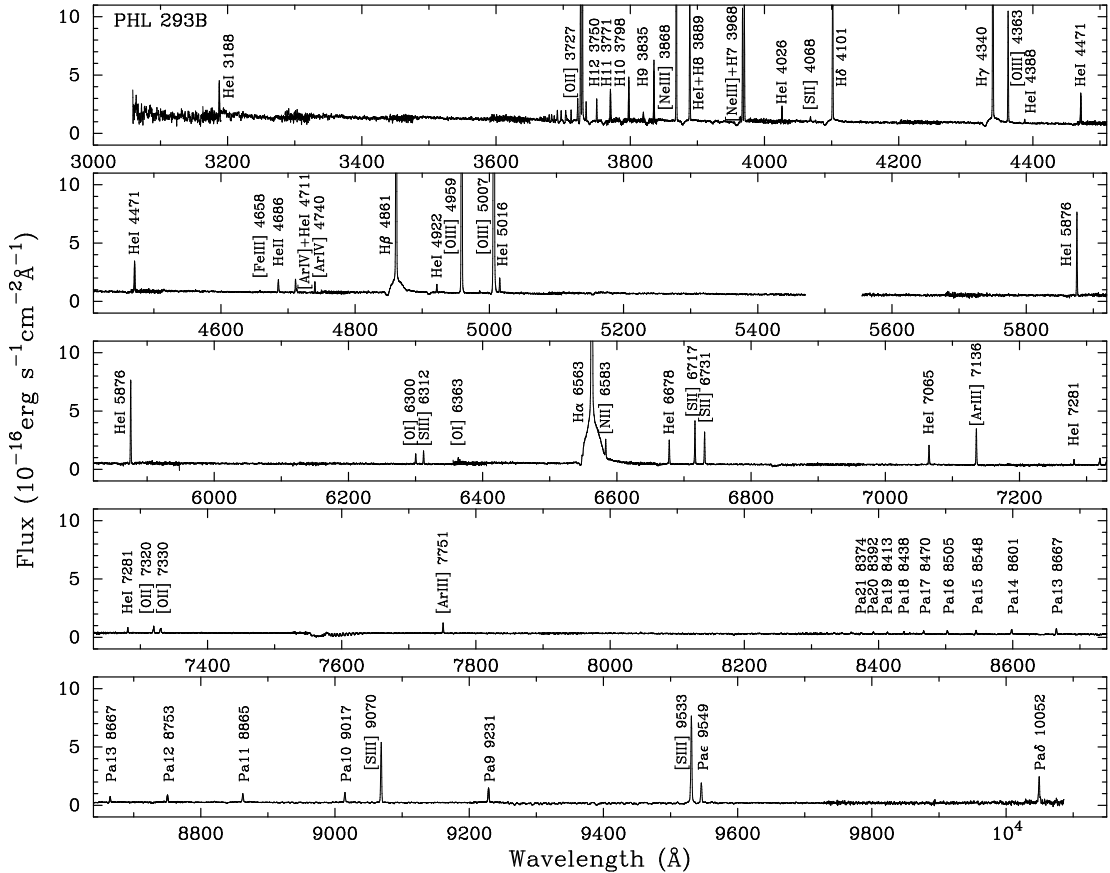,angle=-90,width=14.5cm,clip=}
\caption{Flux-calibrated and redshift-corrected VLT/X-shooter UV and optical 
spectrum of PHL 293B.
Note the broad emission with P Cygni profiles in the hydrogen lines 
H$\delta$ $\lambda$4101 through H$\alpha$ $\lambda$6563.
}
\label{fig1}
\end{figure*}

\begin{figure*}[t]
\hspace*{0.0cm}\psfig{figure=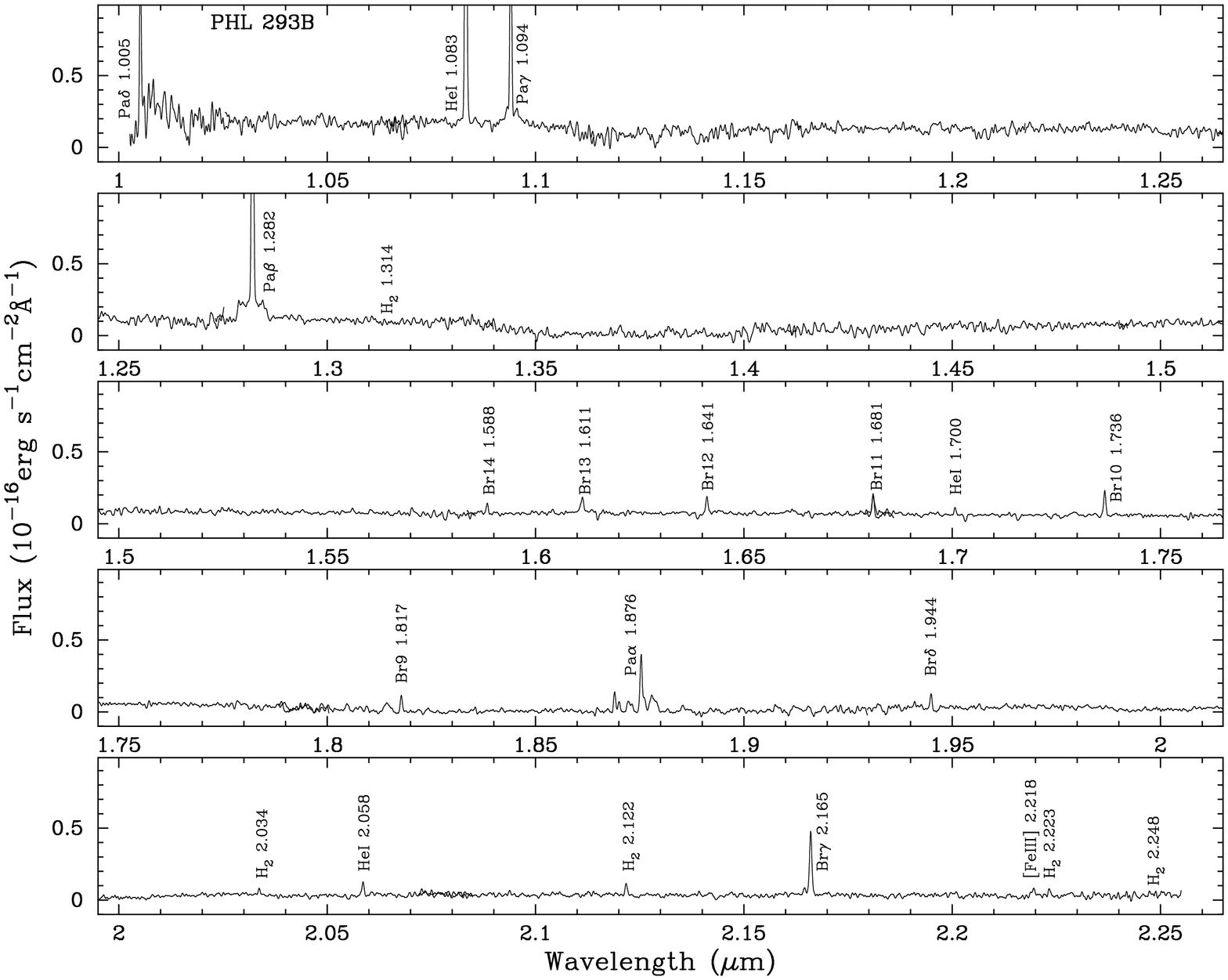,angle=-90,width=14.5cm,clip=}
\caption{The same as in Fig. \ref{fig1} but for the near-infrared (NIR) range.
Note the broad emission in the Pa$\beta$ and Pa$\gamma$ lines. Broad 
emission in the Pa$\alpha$ line is also present but it is strongly absorbed
by telluric lines.
}
\label{fig2}
\end{figure*}

\section{Observations and data reduction \label{obs}}

A new spectrum of PHL 293B was obtained 
using the 8.2 m Very Large Telescope (VLT) on 2009 August 16 
[ESO program 60.A-9442(A)]. The observations were performed 
at an airmass $\sim$ 1.4 using the three-arm echelle X-shooter spectrograph 
mounted at the UT2 Cassegrain focus.
In the UVB arm with wavelength range $\lambda$3000 -- 5595\AA, four equal 
exposures of 900s with a 1\arcsec$\times$11\arcsec\ slit were obtained 
resulting in a total exposure time of 3600s. 
The angular scale of 1\arcsec\ at a distance of 22.7 Mpc to PHL 293B
taken from the NASA/IPAC Extragalactic Database 
(NED)\footnote{NASA/IPAC Extragalactic Database (NED)
is operated by the Jet Propulsion Laboratory, California Institute of 
Technology, under contract with the National Aeronautics and Space 
Administration.} corresponds to a linear scale of 110 pc. 
The distance of PHL 293B has been obtained from the radial velocity
corrected for Virgo Infall with a Hubble constant of 
73 km s$^{-1}$ Mpc$^{-1}$.
Nodding along
the slit was performed according to the scheme ABBA with the object positions
A or B differing by 5\arcsec\ along the slit. In the VIS arm with wavelength 
range $\lambda$5595 -- 10240\AA,
eight equal exposures of 450s with a 0\farcs9$\times$11\arcsec\ slit
were obtained requiring a total exposure time of 3600s. Nodding along
the slit was performed according to the scheme AABBBBAA.
In the NIR arm with wavelength range $\lambda$10240 -- 24800\AA,
twelve equal exposures of 60s with a 0\farcs9$\times$11\arcsec\ slit
were obtained resulting in a total exposure time of 720s. Nodding along
the slit was performed according to the scheme AAABBBBBBAAA.
The above instrumental set-up resulted in resolving powers 
$\lambda$/$\Delta$$\lambda$ of 5100, 8800 and 5100 for the UVB,
VIS and NIR arms, respectively. The seeing was $\sim$ 1\arcsec.

The Kitt Peak IRS spectrophotometric standard star Feige 110 was observed 
on 2009 August 15 with a slit of 5\arcsec$\times$11\arcsec\ for flux
calibration and several bright stars with the same slits as those used
for PHL 293B were observed on 2009 August 15 to correct for
telluric absorption. 
We use the spectrum of Feige 110 for the flux calibration of 
the spectra obtained
with the UVB and VIS arms. 
Since this star was observed with a wide slit, no light losses need to be 
taken into account for the flux calibration.
As for the NIR arm, we use the spectrum of
one of the telluric stars, Hip 15378,
of spectral type A0V with a $V$ magnitude of 7.86 mag, to perform the flux
calibration. Since there is no available absolute spectral distribution for
that star, we adopted the spectral energy distribution of another A0V star,
Vega, with a $V$ magnitude of 0.0 mag and scaled it to the $V$ magnitude of
Hip 15378.
Spectra of thorium-argon (Th-Ar) comparison arcs were 
obtained and used for wavelength calibration of the UVB and VIS arm 
observations. For the wavelength calibration of the NIR spectrum we use night 
sky emission lines.

The two-dimensional UVB and VIS spectra were bias subtracted and 
flat-field corrected using IRAF\footnote{IRAF is 
the Image Reduction and Analysis Facility distributed by the 
National Optical Astronomy Observatory, which is operated by the 
Association of Universities for Research in Astronomy (AURA) under 
cooperative agreement with the National Science Foundation (NSF).}. 
The two-dimensional NIR spectra were corrected for a dark current and divided
by the flat frames to correct for the pixel sensitivity variations.
Cosmic ray hits of all UVB, VIS and NIR spectra were removed using routine 
CRMEDIAN. The remaining
hits were later removed manually after background subtraction.
For each of the UVB, VIS and NIR arms we separately coadded spectra
with the object at the position A and spectra with the object at the position 
B. Then, the coadded spectrum at the position B was subtracted from the
coadded spectrum at the position A. This resulted in a frame with
subtracted background. We used the IRAF
software routines IDENTIFY, REIDENTIFY, FITCOORD, and TRANSFORM to 
perform wavelength
calibration and correct for distortion and tilt of each frame. 
The one-dimensional wavelength-calibrated spectra were then extracted from 
the two-dimensional frames using the APALL routine. 
We adopted extraction apertures of  
1\arcsec$\times$2\farcs5, 0\farcs9$\times$2\farcs5 and 
0\farcs9$\times$2\farcs5 for the UVB, VIS and NIR spectra, respectively.

The resulting flux-calibrated
 and redshift-corrected UVB and VIS spectra of 
PHL 293B are shown in Fig. \ref{fig1} and the resulting flux-calibrated 
and redshift-corrected NIR spectrum is shown in Fig. \ref{fig2}. 
Despite the fact that UVB and VIS spectra, on the one hand, and the NIR 
spectrum, on the other hand, were flux-calibrated with the use of two 
different stars observed with different apertures, the levels of the continuum
at $\sim$ 10000\AA\ in the VIS and NIR spectra agree to better than
10\%. Therefore, no adjustment of the VIS and NIR spectra was applied.
Strong broad hydrogen emission lines with P Cygni profiles are present in the 
UVB and VIS spectra, similar to the ones seen in the UVES spectrum
\citep{IT09}. Two hydrogen lines in the NIR spectrum 
(Fig. \ref{fig2}), Pa$\beta$ and Pa$\gamma$, also show the broad components.


\begin{table*}
\caption{Narrow emission-line fluxes with 1$\sigma$ errors\label{tab1}}
\begin{tabular}{lrr|lrr} \hline\hline
Line                 &100$\times$$F$($\lambda$)/$F$(H$\beta$)$^a$&100$\times$$I$($\lambda$)/$I$(H$\beta$)$^b$&
Line                 &100$\times$$F$($\lambda$)/$F$(H$\beta$)$^a$&100$\times$$I$($\lambda$)/$I$(H$\beta$)$^b$ \\ \hline
3188 He {\sc i}      &  2.98$\pm$0.22&  3.73$\pm$0.35&6717 [S {\sc ii}]    &  5.28$\pm$0.08&  4.45$\pm$0.28  \\  
3683 H20             &  0.55$\pm$0.04&  2.39$\pm$0.28&6731 [S {\sc ii}]    &  3.96$\pm$0.07&  3.33$\pm$0.24  \\
3687 H19             &  0.47$\pm$0.04&  2.29$\pm$0.30&7065 He {\sc i}      &  2.66$\pm$0.05&  2.18$\pm$0.19  \\
3692 H18             &  1.02$\pm$0.04&  2.92$\pm$0.28&7136 [Ar {\sc iii}]  &  4.75$\pm$0.07&  3.89$\pm$0.26  \\
3697 H17             &  0.98$\pm$0.04&  2.90$\pm$0.27&7281 He {\sc i}      &  0.66$\pm$0.02&  0.54$\pm$0.10  \\
3704 H16             &  0.95$\pm$0.05&  2.90$\pm$0.28&7320 [O {\sc ii}]    &  1.07$\pm$0.03&  0.87$\pm$0.12  \\
3705 He {\sc i}      &  0.52$\pm$0.05&  0.61$\pm$0.12&7330 [O {\sc ii}]    &  1.04$\pm$0.03&  0.84$\pm$0.12  \\
3712 H15             &  1.21$\pm$0.04&  2.90$\pm$0.27&7751 [Ar {\sc iii}]  &  1.42$\pm$0.03&  1.11$\pm$0.14  \\
3722 H14             &  2.57$\pm$0.06&  4.56$\pm$0.34&8392 Pa20            &  0.35$\pm$0.02&  0.57$\pm$0.10  \\
3726 [O {\sc ii}]    & 18.42$\pm$0.27& 21.43$\pm$0.74&8413 Pa19            &  0.32$\pm$0.02&  0.53$\pm$0.09  \\
3729 [O {\sc ii}]    & 25.56$\pm$0.38& 29.73$\pm$0.88&8438 Pa18            &  0.40$\pm$0.02&  0.60$\pm$0.10  \\
3734 H13             &  2.02$\pm$0.06&  3.97$\pm$0.31&8446 O {\sc i}       &  0.13$\pm$0.01&  0.10$\pm$0.04  \\
3750 H12             &  2.18$\pm$0.05&  4.23$\pm$0.32&8470 Pa17            &  0.45$\pm$0.02&  0.65$\pm$0.10  \\
3770 H11             &  2.92$\pm$0.06&  4.95$\pm$0.35&8505 Pa16            &  0.64$\pm$0.02&  0.76$\pm$0.11  \\
3797 H10             &  4.08$\pm$0.09&  6.39$\pm$0.40&8548 Pa15            &  0.65$\pm$0.02&  0.77$\pm$0.11  \\
3820 He {\sc i}      &  0.88$\pm$0.07&  1.02$\pm$0.16&8601 Pa14            &  0.90$\pm$0.03&  0.95$\pm$0.12  \\
3835 H9              &  5.91$\pm$0.11&  8.41$\pm$0.45&8667 Pa13            &  0.97$\pm$0.02&  0.99$\pm$0.13  \\
3868 [Ne {\sc iii}]  & 37.69$\pm$0.55& 43.18$\pm$1.08&8753 Pa12            &  1.28$\pm$0.04&  1.23$\pm$0.14  \\
3889 He {\sc i}+H8   & 17.35$\pm$0.26& 21.59$\pm$0.74&8865 Pa11            &  1.71$\pm$0.04&  1.54$\pm$0.16  \\
3964 He {\sc i}      &  0.55$\pm$0.05&  0.63$\pm$0.12&9017 Pa10            &  2.02$\pm$0.04&  1.73$\pm$0.16  \\
3967 [Ne {\sc iii}]  & 11.34$\pm$0.17& 12.83$\pm$0.56&9069 [S {\sc iii}]   & 10.70$\pm$0.16&  8.05$\pm$0.36  \\
3970 H7              & 12.83$\pm$0.20& 16.14$\pm$0.63&9232 Pa9             &  2.75$\pm$0.06&  2.25$\pm$0.19  \\
4026 He {\sc i}      &  1.52$\pm$0.06&  1.71$\pm$0.20&9532 [S {\sc iii}]$^c$& 16.67$\pm$0.24& 12.25$\pm$0.46  \\
4068 [S {\sc ii}]    &  0.54$\pm$0.04&  0.61$\pm$0.12&9549 Pa$\epsilon$    &  3.94$\pm$0.06&  3.09$\pm$0.22  \\
4076 [S {\sc ii}]    &  0.16$\pm$0.03&  0.18$\pm$0.07&10052 Pa$\delta$     &  5.02$\pm$0.13&  3.78$\pm$0.24  \\
4101 H$\delta$       & 22.85$\pm$0.33& 26.84$\pm$0.81&10052 Pa$\delta$(NIR)&  5.75$\pm$0.53&  4.40$\pm$0.43  \\
4340 H$\gamma$       & 42.91$\pm$0.62& 47.39$\pm$1.09&10829 He {\sc i}     & 27.38$\pm$0.49& 18.96$\pm$0.60  \\
4363 [O {\sc iii}]   & 12.36$\pm$0.18& 13.19$\pm$0.55&10941 Pa$\gamma$     & 10.02$\pm$0.26&  7.11$\pm$0.34  \\
4388 He {\sc i}      &  0.45$\pm$0.03&  0.48$\pm$0.10&12821 Pa$\beta$      & 22.32$\pm$0.46& 15.07$\pm$0.53  \\
4471 He {\sc i}      &  3.29$\pm$0.07&  3.45$\pm$0.28&15884 Br14           &  0.49$\pm$0.08&  0.38$\pm$0.08  \\
4658 [Fe {\sc iii}]  &  0.26$\pm$0.03&  0.26$\pm$0.08&16114 Br13           &  1.04$\pm$0.11&  0.74$\pm$0.11  \\
4686 He {\sc ii}     &  1.72$\pm$0.04&  1.75$\pm$0.19&16412 Br12           &  0.83$\pm$0.09&  0.60$\pm$0.10  \\
4711 [Ar {\sc iv}]   &  1.61$\pm$0.04&  1.63$\pm$0.19&16811 Br11           &  0.99$\pm$0.07&  0.70$\pm$0.10  \\
4713 He {\sc i}      &  0.45$\pm$0.04&  0.46$\pm$0.10&17006 He {\sc i}     &  0.40$\pm$0.09&  0.26$\pm$0.07  \\
4740 [Ar {\sc iv}]   &  1.32$\pm$0.04&  1.33$\pm$0.17&17367 Br10           &  1.23$\pm$0.08&  0.83$\pm$0.11  \\
4861 H$\beta$        &100.00$\pm$1.42&100.00$\pm$1.44&18179 Br9$^c$        &  0.80$\pm$0.08&  0.51$\pm$0.09  \\
4921 He {\sc i}      &  1.02$\pm$0.03&  1.00$\pm$0.14&18756 Pa$\alpha$$^c$ &  2.57$\pm$0.10&  1.65$\pm$0.15  \\
4959 [O {\sc iii}]   &188.62$\pm$2.68&184.30$\pm$2.82&19451 Br$\delta$$^c$ &  0.64$\pm$0.10&  0.43$\pm$0.09  \\
4986 [Fe {\sc iii}]  &  0.33$\pm$0.02&  0.32$\pm$0.08&20337 H$_2$ 1-0 S(2) &  0.27$\pm$0.05&  0.17$\pm$0.05  \\
5007 [O {\sc iii}]   &558.58$\pm$7.93&542.49$\pm$7.85&20586 He {\sc i}$^c$ &  0.66$\pm$0.06&  0.42$\pm$0.08  \\
5016 He {\sc i}      &  1.88$\pm$0.04&  1.82$\pm$0.19&21220 H$_2$ 1-0 S(1) &  0.67$\pm$0.06&  0.42$\pm$0.08  \\
5876 He {\sc i}      &  9.79$\pm$0.15&  8.74$\pm$0.41&21661 Br$\gamma$     &  3.62$\pm$0.11&  2.28$\pm$0.18  \\
6300 [O {\sc i}]     &  1.14$\pm$0.03&  0.99$\pm$0.13&22195 [Fe {\sc iii}] &  0.35$\pm$0.09&  0.22$\pm$0.07  \\
6312 [S {\sc iii}]   &  1.47$\pm$0.04&  1.27$\pm$0.15&22230 H$_2$ 1-0 S(0) &  0.34$\pm$0.10&  0.21$\pm$0.07  \\
6363 [O {\sc i}]     &  0.62$\pm$0.07&  0.53$\pm$0.10&$C$(H$\beta$)&\multicolumn{2}{c}{0.225}  \\
6563 H$\alpha$       &333.33$\pm$4.73&284.26$\pm$4.45&EW(H$\beta$)$^d$&\multicolumn{2}{c}{112}  \\
6583 [N {\sc ii}]    &  2.04$\pm$0.05&  1.73$\pm$0.18&$F$(H$\beta$)$^e$&\multicolumn{2}{c}{83.8} \\
6678 He {\sc i}      &  3.02$\pm$0.06&  2.56$\pm$0.21&EW(abs)$^d$&\multicolumn{2}{c}{1.15} \\
\hline \\
\end{tabular}

$^a$ observed flux. \\
$^b$ flux corrected for extinction and underlying stellar absorption. \\
$^c$ affected by telluric absorption. \\ 
$^d$ equivalent width of hydrogen absorption lines in \AA. \\
$^e$ in units 10$^{-16}$ erg s$^{-1}$ cm$^{-2}$. 

\end{table*}



\begin{table}[t]
\caption{Physical conditions and element abundances \label{tab2}}
\begin{tabular}{lc} \hline\hline
Property     &Value         \\ \hline 

$T_e^{dir}$(O {\sc iii}), K                  &16690$\pm$360  \\
$T_e$(O {\sc ii}), K                   &15220$\pm$300  \\
$T_e^{dir}$(O {\sc ii}), K             &15820$\pm$1460 \\
$T_e$(S {\sc iii}), K                  &15550$\pm$300  \\
$T_e^{dir}$(S {\sc iii}), K            &15640$\pm$3230 \\
$N_e$(O {\sc ii}), cm$^{-3}$           &   69$\pm$47    \\
$N_e$(S {\sc ii}), cm$^{-3}$           &   76$\pm$166   \\
$N_e$(Ar {\sc iv}), cm$^{-3}$          & 1680$\pm$530   \\ \\
O$^+$/H$^+$, ($\times$10$^5$)
          &0.45$\pm$0.03  \\
O$^{2+}$/H$^+$, ($\times$10$^5$)       &4.63$\pm$0.25  \\
O$^{3+}$/H$^+$, ($\times$10$^6$)       &1.02$\pm$0.15  \\
O/H, ($\times$10$^5$)                  &5.18$\pm$0.25  \\
12+log O/H                             &7.71$\pm$0.02  \\ \\
N$^{+}$/H$^+$, ($\times$10$^6$)        &0.12$\pm$0.01  \\
$ICF$(N)$^a$                           &     10.8      \\
N/H, ($\times$10$^6$)                  &1.34$\pm$0.12  \\
log N/O                                &--1.59$\pm$0.05~~\\ \\
Ne$^{2+}$/H$^+$, ($\times$10$^5$)      &0.86$\pm$0.05  \\
$ICF$(Ne)$^a$                          &      1.04     \\
Ne/H, ($\times$10$^5$)                 &0.89$\pm$0.06  \\
log Ne/O                               &--0.77$\pm$0.03~~\\ \\
S$^{+}$/H$^+$, ($\times$10$^6$)        &0.07$\pm$0.01  \\
S$^{2+}$/H$^+$, ($\times$10$^6$)       &0.59$\pm$0.08  \\
$ICF$(S)$^a$                           &      2.30     \\
S/H, ($\times$10$^6$)                  &1.52$\pm$0.17  \\
log S/O                                &--1.53$\pm$0.05~~\\ \\
Ar$^{2+}$/H$^+$, ($\times$10$^7$)      &1.47$\pm$0.10  \\
Ar$^{3+}$/H$^+$, ($\times$10$^7$)      &1.03$\pm$0.13  \\
$ICF$(Ar)$^a$                          &      2.22     \\
Ar/H, ($\times$10$^7$)                 &3.25$\pm$0.38  \\
log Ar/O                               &--2.20$\pm$0.06~~\\ \\
Fe$^{2+}$/H$^+$, ($\times$10$^6$)(4658)&0.05$\pm$0.01  \\
$ICF$(Fe)$^a$                          &      16.0     \\
Fe/H, ($\times$10$^6$)(4658)           &0.76$\pm$0.22  \\
log Fe/O (4658)                        &--1.83$\pm$0.13~~\\ \\
Fe$^{2+}$/H$^+$, ($\times$10$^6$)(4986)&0.06$\pm$0.01  \\
$ICF$(Fe)$^a$                          &      16.0     \\
Fe/H, ($\times$10$^6$)(4986)           &0.93$\pm$0.24  \\
log Fe/O (4986)                        &--1.75$\pm$0.11~~\\
\hline
\end{tabular}

\smallskip

$^a$ Ionisation correction factor.
\end{table}


\section{The properties of the host galaxy \label{host}}

\subsection{Element abundances \label{abund}}

We derived element abundances from the narrow emission-line fluxes, 
using a classical semi-empirical method. These lines trace the 
interstellar medium (ISM) of PHL 293B. The fluxes in all spectra were 
measured using Gaussian fitting with the IRAF SPLOT routine. 
The 1$\sigma$ errors of the line fluxes were calculated from the
data in the non-flux-calibrated spectra using SPLOT. The method is based on
Monte-Carlo simulations, in which random gaussian 
noise with the dispersion obtained from the continuum near the line
is added to the noise-free spectrum. To obtain good error estimates
we have run 200 simulations per emission line. These errors were
propagated in the calculations of the elemental abundance errors.
The fluxes were corrected for both extinction, using the reddening curve
of \citet{C89}, and underlying
hydrogen stellar absorption, derived simultaneously by an iterative procedure 
described by \citet{ITL94} and using the observed decrements of the 
narrow hydrogen Balmer lines in the UVB and VIS spectra. 
The extinction coefficient 
$C$(H$\beta$) and equivalent width of hydrogen absorption lines EW(abs) are 
derived in such a way as to obtain the closest agreement between the 
extinction-corrected and theoretical recombination 
Balmer hydrogen emission-line 
fluxes normalised to the H$\beta$ flux. It is assumed that EW(abs) is the same
for all hydrogen lines. This assumption is justified by the evolutionary 
stellar population synthesis models of \citet{GD05}. The derived $C$(H$\beta$)
and EW$_{abs}$ are applied for correction of the emission-line fluxes in the
entire wavelength range $\lambda$$\lambda$ 3000 -- 22650\AA.

The extinction-corrected total fluxes 
100$\times$$I$($\lambda$)/$I$(H$\beta$) of the narrow lines, 
and the extinction coefficient
$C$(H$\beta$), the equivalent width of the H$\beta$ emission line 
EW(H$\beta$), the H$\beta$ observed flux $F$(H$\beta$), and the 
equivalent width of the 
underlying hydrogen absorption lines EW(abs) are given in Table \ref{tab1}.
The extinction-corrected fluxes are generally in
good agreement with those derived by \citet{IT09}.

The physical conditions, and the ionic and total heavy element 
abundances in the H~{\sc ii} region of PHL 293B were derived 
following \citet{I06a} (Table \ref{tab2}). 
 In particular for  
O$^{2+}$, Ne$^{2+}$, and Ar$^{3+}$, we adopt
the temperature $T_e^{dir}$(O~{\sc iii}) directly derived from the 
[O~{\sc iii}] $\lambda$4363/($\lambda$4959 + $\lambda$5007)
emission-line ratio.
The electron temperatures $T_e$(O~{\sc ii}) and
$T_e$(S {\sc iii}) were derived from the empirical relations by
\citet{I06a} based on photoionisation models of H {\sc ii} regions.
By direct methods we derived also $T_e^{dir}$(O~{\sc ii})= 15820$\pm$1460 K 
from the [O~{\sc ii}]$\lambda$(3726+3729)/$\lambda$(7320+7330) 
emission-line ratio 
and $T_e^{dir}$(S~{\sc iii})= 15640$\pm$3230 K from the 
[S~{\sc iii}]$\lambda$6312/$\lambda$9069 emission-line ratio.
Despite their higher errors these temperatures are in agreement with 
the temperatures $T_e$(O~{\sc ii}) and $T_e$(S {\sc iii}) 
derived from the empirical relations.
We used $T_e$(O~{\sc ii}) for the calculation of
O$^{+}$,  N$^{+}$, S$^{+}$, and Fe$^{2+}$ abundances and $T_e$(S {\sc iii})
for the calculation of S$^{2+}$ and Ar$^{2+}$ abundances.
The electron number densities  $N_e$(O~{\sc ii}), 
$N_e$(S~{\sc ii}) and $N_e$(Ar {\sc iv}) were obtained from the [O~{\sc ii}] 
$\lambda$3726/$\lambda$3729, [S~{\sc ii}] 
$\lambda$6717/$\lambda$6731 and [Ar~{\sc iv}] 
$\lambda$4711/$\lambda$4740 emission-line ratios, respectively.
For these low electron number densities (Table \ref{tab2}), 
the collisional deactivation 
of the upper levels of forbidden optical and NIR emission lines is negligible 
as compared to the spontaneous transitions. The element abundances then do not 
depend sensitively on $N_e$ and
the low-density limit holds for the abundance determination in PHL 293B.
The ionic and total O, N, Ne,
 S, Ar, and Fe abundances derived from the
forbidden emission lines are given in Table \ref{tab2}.
The oxygen abundance 12+log O/H = 7.71$\pm$0.02
and abundance ratios log N/O = --1.59$\pm$0.05 and 
log Ne/O = --0.77$\pm$0.03
are consistent with the respective values 7.72$\pm$0.01, 
--1.55$\pm$0.02 and --0.74$\pm$0.02 derived by \citet{IT09}. 
All abundance ratios in Table \ref{tab2} are typical for the 
low-metallicity emission-line galaxies \citep{I06a}.

\begin{table} [t]
\caption{H$_2$ emission-line fluxes relative to the H$_2$ 2.122$\mu$m flux 
\label{tab3}}
\begin{tabular}{lccc} \hline \hline
Line&Observations&\multicolumn{2}{c}{Model$^a$} \\ \cline{3-4}
    &            &Fluor.&Collis. \\ \hline 
1.314 H$_2$ 3-1 Q(1)&  $<$0.2   &0.6&0.0 \\
2.034 H$_2$ 1-0 S(2)&0.4$\pm$0.2&0.5&0.3 \\
2.122 H$_2$ 1-0 S(1)&1.0&1.0&1.0 \\
2.223 H$_2$ 1-0 S(0)&0.4$\pm$0.2&0.6&0.3 \\
2.248 H$_2$ 2-1 S(1)&  $<$0.3   &0.5&0.0 \\ \hline
\end{tabular}

$^a$ Model values are from \citet{BD87}. We adopt their
model 1 for fluorescent lines and model S1 for collisionally excited lines.
\end{table}

\subsection{Extinction and hidden star formation\label{sec:EXT}}

In all previous studies of BCDs except for those of Mrk 59 in 
\citet{I09}, II Zw 40, Mrk 71, Mrk 996 and SBS 0335--052E in
\citet{IT11}, the NIR spectra of BCDs were obtained in separate $JHK$
observations, and there was no wavelength overlap between the 
optical and NIR spectra. The fact that the spectrum of PHL 293B has been 
obtained simultaneously over the entire optical and near-infrared wavelength  
ranges, avoids adjusting uncertainties and 
permits us to compare directly the optical and NIR extinctions.

The agreement between the corrected values of the hydrogen line fluxes in 
both the optical and NIR 
ranges (Table \ref{tab1}) and theoretical recombination values 
by \citet{HS87} implies that a 
single $C$(H$\beta$) [or $A(V)$] can be used over the whole 
3000 -- 22650\AA\ range to correct line fluxes for extinction.
In particular, the extinction-corrected flux of the Br$\gamma$ emission
line $I$(Br$\gamma$)/$I$(H$\beta$) = 0.23$\pm$0.02 is in agreement with
the \citet{HS87} recombination value of 0.25 for the electron temperature 
$T_e$ = 15000K and the electron number density $N_e$ = 100 cm$^{-3}$. 
The fact that $A(V)$ does not increase when going from the optical to the 
NIR wavelength ranges implies that the NIR emission lines do not probe more 
extinct regions with hidden star formation as compared to the optical emission 
lines. This appears to be a general result for BCDs 
\citep[e.g.][]{V00,V02,I09,IT11}. 

\subsection{H$_2$ emission\label{sec:H2}}

Molecular hydrogen lines do not originate in the H {\sc ii} region, but in 
neutral molecular clouds. In the near-infrared, they are excited through two
mechanisms. The first one is a thermal mechanism consisting of 
collisions between neutral species (e.g., H, H$_2$) resulting from large-scale 
shocks due to the interaction of stars with molecular clouds or/and
cloud-cloud collisions. The second one is the
fluorescent mechanism due to the absorption of ultraviolet photons
from the hot stars.
By comparing the observed line ratios with 
those predicted by models, such as those calculated by \citet{BD87}, it is 
possible to discriminate between the two processes. 
In particular, line emission from vibrational levels $v$$\geq$2 
is virtually absent in collisionally excited spectra, while they are 
relatively strong in fluorescent spectra.

Three H$_2$ lines,
2.034 $\mu$m 1-0 S(2), 2.122 $\mu$m 1-0 S(1) and 2.223 $\mu$m 1-0 S(0),
are detected in the NIR spectrum of PHL 293B. 
These lines are labelled in Fig. \ref{fig2} and their fluxes relative 
to that of the strongest
2.122 $\mu$m 1-0 S(1) line with the 1$\sigma$ errors are shown in 
Table \ref{tab3}. 
The two last columns of the Table show theoretical ratios by \citet{BD87} 
for fluorescent and collisional excitation, respectively.
It is seen from the Table that the observed line flux ratios of the lines
from the vibrational level $v$ = 1 are
in agreement with the values expected for both the fluorescent 
and collisional excitation. However, no emission lines from the vibrational 
levels $v$ $\geq$ 2 were detected while they should be
present if the fluorescent excitation is important. 

We show in Table \ref{tab3} the 1$\sigma$ upper limits of the two 
strongest H$_2$ emission lines, 1.314 $\mu$m 3-1 Q(1) and 
2.248 $\mu$m 2-1 S(1), originating from the upper levels $v$ $\geq$2 and 
label their location in the spectrum on Fig. \ref{fig2}. In the case of
excitation dominated by fluorescence these lines would be as strong as 
2.034 $\mu$m 1-0 S(2) and 2.223 $\mu$m 1-0 S(0) lines.
Therefore, we conclude
that H$_2$ vibrational states in the spectrum of PHL 293B are mainly
collisionally excited. This is at variance with conclusions made by 
\citet{V00} for SBS 0335--052E, \citet{V08} for II Zw 40, \citet{I09} 
for Mrk 59, and \citet{IT11} for II Zw 40, Mrk 71, Mrk 930, Mrk 996 and 
SBS 0335--052E. There it has been found that fluorescence 
is the dominating excitation mechanism of H$_2$ lines in BCDs.
Apparently, the difference is related to the low H$\beta$ luminosity of
PHL293B that is $\sim$ 6 times lower than that in Mrk 71 and more than one
order of magnitude lower than that in other BCDs studied by \citet{V00}, 
\citet{V08}, \citet{I09} and \citet{IT11}, indicating a low intensity of the
UV radiation in PHL293B.

\section{Broad emission from the LBV star \label{lbv}}

The detection of the LBV star in PHL 293B indicates that high-mass
stars are present in its star forming region. 
The equivalent width of the narrow H$\beta$ emission line
EW(H$\beta$) = 112\AA\ in PHL 293B
(Table \ref{tab1}) corresponds to a
starburst age of $\sim$ 4.5 Myr for the PHL 293B metallicity \citep{SV98}. 
At this age massive stars with masses up to $\sim$ 60 $M_\odot$ still can be 
present \citep{M94} and one of them was likely evolved to the LBV star
we see today.
Below we consider the properties of the LBV broad emission lines seen in 
the VLT/X-shooter spectrum.

\subsection{Fluxes\label{sec:broadfluxes}}

\begin{figure}[t]
\hspace*{0.0cm}\psfig{figure=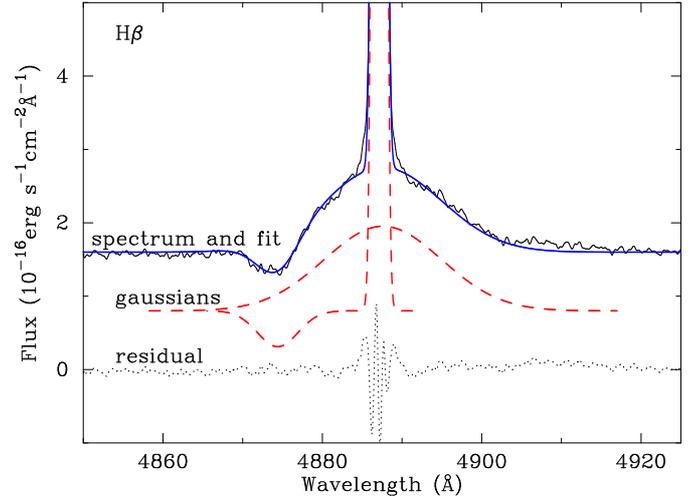,angle=-90,width=8.8cm,clip=}
\caption{ Decomposition of the H$\beta$ profile into absorption
and narrow and broad emission gaussian profiles. The observed
spectrum and the fit to it are shown by solid lines, gaussian
profiles are shown by dashed lines, and the residual spectrum
is shown by the dotted line. For a better view, the spectrum, the fit and 
gaussians are shifted up by 
8$\times$10$^{-17}$ erg s$^{-1}$cm$^{-2}$\AA$^{-1}$.
}
\label{fig3}
\end{figure}

In Table \ref{tab4} we show the observed fluxes of the broad 
and narrow components of hydrogen H$\gamma$, H$\beta$ and 
H$\alpha$ emission lines for three epochs encompassing
$\sim$ 8 years. 
For the SDSS and UVES data, we adopted fluxes and respective
errors from \citet{IT09}. For the X-shooter data, 
using the IRAF/SPLOT routine, we decomposed the H$\gamma$, 
H$\beta$ and H$\alpha$ lines into a narrow emission gaussian component, 
a broad emission gaussian 
component and an absorption gaussian component. An example of the line 
decomposition
is shown in Fig. \ref{fig3} for the H$\beta$ line profile. It is seen
from the Figure that the fit nicely reproduces the observed line profile. 
Fluxes of the broad
lines in Table \ref{tab4} are those corresponding to the fluxes
of broad gaussians with the respective errors which are derived 
similarly to those for narrow lines according to prescriptions in 
Sect. \ref{abund}. We note that real errors of the fluxes 
can be significantly larger due to, e.g., uncertainties 
of the flat-field correction, background subtraction, telluric absorption,
flux calibration, which are difficult to be accounted for.

We also note for the first time the presence of broad components
of Pa$\gamma$ and Pa$\beta$ emission lines in the VLT/X-shooter spectrum
with the observed fluxes of (4.7$\pm$0.7)$\times$10$^{-16}$ and
(8.1$\pm$0.5)$\times$10$^{-16}$ erg s$^{-1}$ cm$^{-2}$, respectively.

A comparison of fluxes in Table \ref{tab4} for 
different epochs shows that they vary by $\la$ 50\%
and are the lowest for the VLT/UVES 
observations for both the broad and narrow components. The fluxes
of broad components in SDSS and VLT/X-shooter spectra are in
agreement, but fluxes of narrow components are higher in the SDSS
spectrum. The narrow lines are most likely produced in the H {\sc ii}
region ionised by many O stars. Therefore, we would not expect temporal
variations of these lines. Probably, the
observational uncertainties and different extraction apertures
play a role. The observational uncertainties may be caused by varying 
weather conditions, uncertainties in acquisition, and
different epochs of observation of the object and the standard star.
The larger fluxes of narrow emission lines by $\sim$25\% 
in the SDSS spectrum can be explained by a larger round 3\arcsec\ aperture
as compared to the smaller extraction aperture for the
VLT/X-shooter spectrum. The different apertures, however, should not
influence the fluxes of the broad lines because these lines are emitted by the
compact expanding envelope around the LBV star. As for the VLT/UVES 
observations, \citet{IT09} do not present extraction apertures. 
These observations were done under non-photometric conditions, according to 
the ambient conditions database
of the Paranal observatory. This may be the reason why the fluxes
of the emission lines in the VLT/UVES spectrum are the lowest.

\begin{table*}[t]
\caption{Observed fluxes of the broad and narrow hydrogen emission lines at different epochs 
\label{tab4}}
\begin{tabular}{lccccrrrcccc} \hline\hline
Date&\multicolumn{3}{c}{Broad line flux$^a$}&&\multicolumn{3}{c}{Narrow line flux$^a$} 
&&\multicolumn{3}{c}{Broad-to-narrow line flux ratio} \\ \cline{2-4} \cline{6-8} \cline{10-12}
    &H$\gamma$&H$\beta$
&H$\alpha$&&\multicolumn{1}{c}{H$\gamma$}&\multicolumn{1}{c}{H$\beta$}&\multicolumn{1}{c}{H$\alpha$}   
&&\multicolumn{1}{c}{H$\gamma$}&\multicolumn{1}{c}{H$\beta$}&\multicolumn{1}{c}{H$\alpha$} \\ \hline
2001.08.22$^b$      & ...            &26.6$\pm$4.5 & 97.8$\pm$6.3       && 46.1$\pm$2.1&104.8$\pm$4.6 &351.0$\pm$7.0&& ...         &0.25$\pm$0.04&0.28$\pm$0.02 \\
2002.11.08$^c$  & 7.5$\pm$0.2    &16.7$\pm$0.3 & 55.2$\pm$0.6       && 32.6$\pm$0.5& 71.6$\pm$0.8 &213.6$\pm$2.1&&0.23$\pm$0.01&0.23$\pm$0.01&0.26$\pm$0.01 \\
2009.08.16$^d$      & 7.4$\pm$0.2    &22.8$\pm$0.3 & 97.0$\pm$0.7       && 36.0$\pm$0.5& 83.8$\pm$1.2 &279.3$\pm$4.0&&0.21$\pm$0.01&0.27$\pm$0.01&0.35$\pm$0.01 \\
\hline \\
\end{tabular}

$^a$in units of 10$^{-16}$ erg s$^{-1}$ cm$^{-2}$.

$^b$fluxes are from the SDSS spectrum obtained with the round 
aperture of 3\arcsec\ in diameter \citep{IT09}.

$^c$fluxes from the VLT/UVES spectrum obtained with a 1\arcsec\ wide slit 
under non-photometric conditions \citep{IT09}.

$^d$fluxes from the VLT/Xshooter spectra within the apertures 
1\arcsec$\times$2\farcs5 for the H$\gamma$ and H$\beta$ lines and 
0\farcs$9\times$2\farcs5 for the H$\alpha$ line (this paper).
\end{table*}

We also present in Table \ref{tab4} the broad-to-narrow line flux ratios.
At variance to the line fluxes these flux ratios should not be sensitive to 
the uncertainties in the flux calibration, although they depend on the 
extraction apertures. The largest differences of the flux ratios are for 
the H$\alpha$
line with the largest value in the X-shooter observations. If, however,
the aperture correction by a factor of $\sim$1.25 to the narrow H$\alpha$ line
flux in the X-shooter observations is applied, the broad-to-narrow H$\alpha$ 
line flux ratios in the SDSS and X-shooter data would be in good agreement.
The same aperture correction for the UVES data results in a H$\alpha$
flux ratio, which is $\sim$40\% lower. Apparently, some variations of the
broad emission-line fluxes in PHL 293B on a time scale of 
$\sim$ 8 years were present. These temporal variations are followed by
variations of the broad line widths, discussed below in 
Sect. \ref{sec:broadvel}.

The broad H$\alpha$ line luminosity of the LBV in PHL 293B 
derived from the X-shooter spectrum and corrected 
for the extinction with $C$(H$\beta$) = 0.225 is
9.0 $\times$ 10$^{38}$ erg s$^{-1}$. It compares well with the broad 
H$\alpha$
luminosity of $\sim$ 10$^{39}$ erg s$^{-1}$ derived from the SDSS spectrum
adopting $C$(H$\beta$) = 0.245 from \citet{I07}. Note that the broad H$\alpha$ 
luminosity of the LBV in PHL 293B of 6.0 $\times$ 10$^{38}$ erg s$^{-1}$,
reported by \citet{IT09}, was not corrected for extinction.
Thus, the H$\alpha$ luminosity of the LBV in PHL 293B is 
one order of magnitude higher than 
$L$(H$\alpha$) $\sim$ 10$^{38}$ erg~s$^{-1}$ of the V1 star in NGC 2363,
the third lowest-metallicity LBV known \citep{D01} after those   
in PHL 293B and DDO 68, and two orders of magnitude higher than 
$L$(H$\alpha$) $\sim$ 9.4$\times$10$^{36}$ erg~s$^{-1}$ of the LBV in
DDO 68.
We also can estimate the brightness of the LBV in the continuum.
The intensity of, e.g., H$\gamma$ and H$\beta$ at the 
bottom of the blue absorption features 
in the VLT and SDSS spectra of PHL 293B is
0.8 that of the continuum. Since the light of the LBV star is diluted
by the emission of the galaxy and assuming that the intensities at the bottom 
of these absorption features in the spectrum of the LBV itself are zero 
\citep[e.g., ][]{KP00}, 
we adopt that the fraction of the LBV continuum in the total continuum of 
PHL 293B is $\sim$ 20\%. 
With this fraction and using the apparent SDSS $g$
magnitude of 18.2 inside the 3\arcsec\ fiber aperture of the SDSS spectrum we
obtain an apparent magnitude for the LBV of $g$ = 19.9, which is not corrected 
for extincion. This corresponds to an
absolute magnitude $M_g$ = --11.9 at a distance $D$ = 22.7 Mpc. 
If the extinction coefficient $C$ = 0.225 is taken into account, then the 
LBV star would be $\sim$ 0.5 mag brighter. Thus, despite these crude
estimates, we conclude, that the absolute
brightness of the LBV in PHL 293B is one of the highest among the other LBVs
and compares well with the brightness of $\eta$ Car, the most luminous LBV in 
the Milky Way \citep{M03,S10}. All this implies that the LBV star in PHL 293B 
is experiencing a strong outburst.

\subsection{Velocities of the stellar wind\label{sec:broadvel}}

\begin{figure*}[t]
\hspace*{0.0cm}\psfig{figure=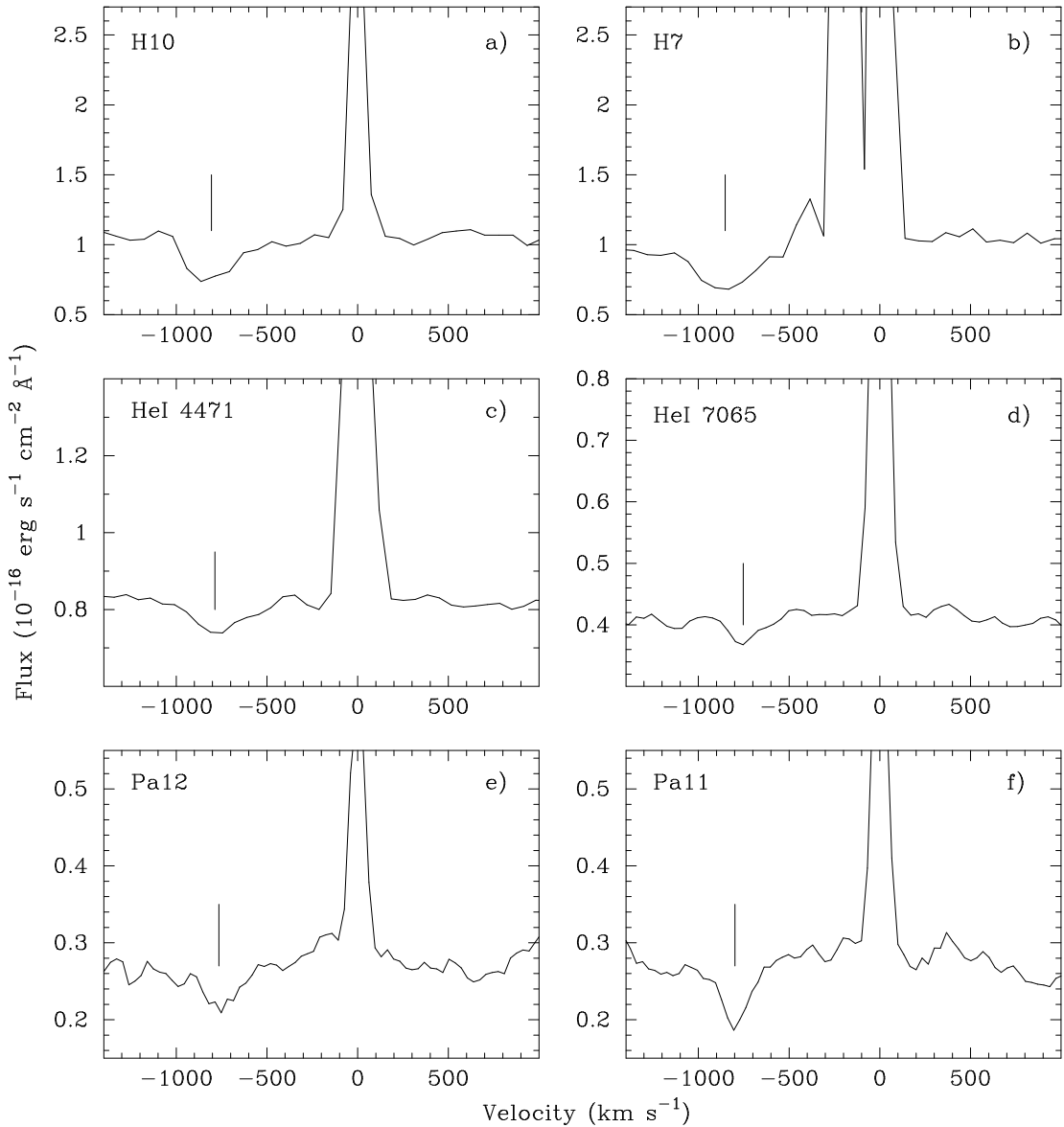,angle=-90,width=14.5cm,clip=}
\caption{The profiles of some Balmer hydrogen, Paschen hydrogen and 
He {\sc i} lines in the VLT/X-shooter spectrum showing the blue-shifted absorption 
(marked by vertical tics).}
\label{fig4}
\end{figure*}

One of the most important features of the VLT/X-shooter spectrum is that
the P Cygni profiles of the LBV star are seen not only in the Balmer hydrogen 
lines, but also in the Paschen hydrogen and He {\sc i} lines. 
In particular, no He {\sc i} lines with P Cygni profiles were detected and reported
in the SDSS and VLT/UVES spectra of PHL 293B implying 
some changes in the LBV spectrum after eight years.
Examples of the lines with the blue absorption minima are shown in Fig. \ref{fig4}.
Earlier \citet{IT09} reported the presence of He {\sc i} lines with P Cygni 
profiles in the spectrum 
of another LBV in the extremely metal-deficient BCD DDO 68.

In Table \ref{tab5}, we show the terminal velocities $v_t$ of the stellar 
winds for the Balmer, Paschen and He {\sc i} lines
and the FWHMs of the broad Balmer and Paschen lines of the PHL 293B 
LBV, as derived from the X-shooter (this paper) and UVES \citep{IT09} 
spectra. The stellar wind terminal velocity $v_t$ is 
derived from the wavelength
difference between the blue absorption minimum and the
narrow nebular emission line 
maximum. The $v_t$'s derived from different lines for two different epochs
are similar. However, the much larger number of the lines of different
species in the VLT/X-shooter spectrum allowed us to uncover some new 
interesting properties. The examination of Table \ref{tab5} shows that the 
terminal velocities derived from the Balmer lines are systematically higher 
than those derived from the Paschen lines and He {\sc i} lines. Thus, the 
averaged terminal velocities for Balmer, Paschen and He {\sc i} lines are 
818$\pm$20, 782$\pm$13 and 762$\pm$34 km s$^{-1}$, respectively,
while the weighted mean terminal velocities for the same lines are
820$\pm$8, 783$\pm$22 and 764$\pm$15 km s$^{-1}$, respectively. 
These differences may have important implications for our understanding of the 
LBV wind properties. The Balmer absorption lines are formed in transitions 
from the level with the principal quantum number $n$ = 2, while for the
Paschen line transition the lower level is $n$ = 3, which is less populated.
Therefore, the Balmer absorption lines are formed in more outer layers of the 
expanding envelope as compared to the Paschen absorption lines. 
Probably, the level $n$ = 2 for He {\sc i} is also less populated as compared
to that for hydrogen because of its higher energy and lower He abundance.
Therefore, the He {\sc i} absorption lines are formed in more inner layers
as compared to the Balmer absorption lines.
Then the differences in the terminal velocities for different
transitions can be explained by a wind accelerating outward.


\begin{table}[t]
\caption{Terminal velocities and FWHMs of the broad emission
lines in PHL 293B at different epochs \label{tab5}}
\begin{tabular}{lccccc} \hline\hline
    & \multicolumn{2}{c}{X-shooter$^a$}&&\multicolumn{2}{c}{UVES$^b$} \\ \cline{2-3} \cline{5-6}
Line&$v_{t}$$^c$&FWHM$_{br}$$^c$&&$v_{t}$$^c$&FWHM$_{br}$$^c$ \\ \hline
3750 H12         &  833$\pm$26     &      ...        &&   ...   &   ...   \\
3770 H11         &  828$\pm$26     &      ...        &&   ...   &   ...   \\
3797 H10         &  806$\pm$24     &      ...        &&   ...   &   ...   \\
3835 H9          &  801$\pm$27     &      ...        &&   ...   &   ...   \\
3888 H8          &  816$\pm$24     &      ...        &&   ...   &   ...   \\
3968 H7          &  848$\pm$24     &      ...        &&   ...   &   ...   \\
4101 H$\delta$   &  792$\pm$24     &  878$\pm$40     && 735     & 474$\pm$30 \\
4340 H$\gamma$   &  836$\pm$24     & 1006$\pm$20     && 871     & 681$\pm$14 \\
4471 He {\sc i}  &  787$\pm$43     &      ...        &&   ...   &   ...   \\
4861 H$\beta$    &  833$\pm$25     &  966$\pm$16     && 819     & 649$\pm$12 \\
4921 He {\sc i}  &  697$\pm$39     &      ...        &&   ...   &   ...   \\
5016 He {\sc i}  &  802$\pm$30     &      ...        &&   ...   &   ...   \\
5876 He {\sc i}  &  777$\pm$55     &      ...        &&   ...   &   ...   \\
6563 H$\alpha$   &  787$\pm$33     &  914$\pm$10     && 859     & 719$\pm$10 \\
6678 He {\sc i}  &  757$\pm$27     &      ...        &&   ...   &   ...   \\
7065 He {\sc i}  &  752$\pm$43     &      ...        &&   ...   &   ...   \\
8753 Pa12        &  765$\pm$42     &      ...        &&   ...   &   ...   \\
8865 Pa11        &  799$\pm$37     &      ...        &&   ...   &   ...   \\
9017 Pa10        &  788$\pm$54     &      ...        &&   ...   &   ...   \\
9232 Pa9         &  775$\pm$53     &      ...        &&   ...   &   ...   \\
10941 Pa$\gamma$ &  ...     & 1237$\pm$160    && ...     &   ...     \\
12821 Pa$\beta$  &  ...     & 1093$\pm$55     && ...     &   ...     \\ \hline
\end{tabular}

$^a$This paper. \\
$^b$\citet{IT09}. \\
$^c$In km s$^{-1}$.
\end{table}

Another important feature of the VLT/X-shooter spectrum is that 
our derived FWHMs of $\sim$ 1000 km s$^{-1}$ are 
by a factor of $\sim$ 1.5 higher than those obtained by \citet{IT09}, but are
consistent with FWHMs derived by \citet{IT09} for another low-metallicity
LBV in the BCD DDO 68. 
We also measured $v_t$'s and FWHMs of broad H$\beta$ and H$\alpha$ emission
lines in the lower-resolution SDSS spectrum obtained $\sim$ 1 year earlier 
than the VLT/UVES spectrum. We find that a $v_t$ of $\sim$ 800 km s$^{-1}$ 
in the SDSS spectrum is
consistent with the values in Table \ref{tab5}. On the other hand, the
FWHMs of the broad H$\beta$ and H$\alpha$ emission lines in the SDSS spectrum,
$\sim$ 1000 km s$^{-1}$, are higher than those in the VLT/UVES spectrum
but are similar to the values derived from the VLT/X-shooter spectrum.
Despite these differences, all FWHMs of the broad lines in the spectra of
PHL 293B are significantly broader than those observed in the spectra of 
high-metallicity LBVs which are about 100-200 km s$^{-1}$ \citep[e.g. ][]{L94}.
However, they are similar to the FWHMs in some extragalactic LBVs during
their outburst phase \citep{S10}.
It is possible therefore that the LBV in PHL 293B also experiences
a strong outburst.

\section{Conclusions \label{concl}}

We have studied here the properties of the narrow-line spectrum 
of the blue compact dwarf (BCD) galaxy PHL 293B and the broad 
line emission of the luminous blue variable (LBV) star in this galaxy 
from the VLT/X-shooter spectroscopic observations in the 
wavelength range $\lambda$$\lambda$3000 -- 22650\AA. This data has been 
compared with the data obtained earlier by \citet{IT09} from SDSS and
VLT/UVES observations. We have arrived at the following conclusions:

1. The oxygen abundance in  PHL 293B is
12+log O/H = 7.71 $\pm$ 0.02, in agreement with the 
12+log O/H = 7.72 $\pm$ 0.01 derived by \citet{IT09} from the
VLT/UVES spectrum and the 12+log O/H = 7.66 $\pm$ 0.04 derived by \citet{I07}
from a lower resolution SDSS spectrum. 
This BCD is the second lowest-metallicity galaxy with a detected LBV star,
after DDO 68. 

2. We find that the extinction-corrected fluxes of narrow hydrogen lines in 
the entire wavelength range $\lambda$$\lambda$ 3000 -- 22650\AA\ 
are in good agreement with the theoretical recombination values if
the extinction coefficient $C$(H$\beta$) = 0.225, derived from the
narrow-line Balmer decrement, is adopted (Table \ref{tab1}). 
This implies that no additional star formation is found that is seen in 
the NIR range but is hidden in the visible range.

3. We detect three $v$ = 1-0 emission lines of vibrationally-excited 
molecular hydrogen. The relative fluxes of these lines 
and non-detection of $v$ = 2-1 and 3-1 emission suggest that H$_2$ is 
collisionally excited in PHL 293B.

4. Broad hydrogen emission lines are seen in both the optical
and NIR spectra of PHL 293B.
The broad H$\alpha$ luminosity in the LBV is by a factor of $\sim$ 10
higher than the one in the higher-metallicity LBV in NGC 2363
\citep{D97,D01,P06} and by a factor of $\sim$ 100 higher than in
the LBV star in DDO 68 \citep{IT09}. We 
find some evidence for flux variations of the broad 
lines on a time scale of $\sim$ 8 years.

5. For the first time we find P Cygni profiles not only
from Balmer hydrogen lines, but also from Paschen hydrogen and He {\sc i} lines
in this LBV.
We derive a high terminal velocity $v_t$ of the stellar wind.
The averaged terminal velocity from Balmer lines 
(818$\pm$20 km s$^{-1}$) is larger than those derived from the Paschen lines
(782$\pm$13 km s$^{-1}$) and He {\sc i} lines (762$\pm$34 km s$^{-1}$), 
probably indicating that the wind is accelerating outward.
These $v_t$'s are
several times larger than the $v_t$ of $\sim$ 100 -- 200 km s$^{-1}$ in 
some high-metallicity counterparts of the Milky Way. However, the high
$v_t$ is similar
to those observed in some extragalactic LBVs during their outburst phase
\citep{S10}.

\acknowledgements
We thank Dr. G. Stasi\'nska for carefully reading the manuscript and 
for providing useful comments.
Y.I.I., N.G.G. and K.J.F. are grateful to the staff of the Max Planck 
Institute for Radioastronomy for their warm hospitality and 
acknowledge support through DFG 
grant No. FR 325/59-1. 
This research has made use of the 
NASA/IPAC Extragalactic Database (NED) which is operated by the Jet Propulsion 
Laboratory, California Institute of Technology, under contract with the 
National Aeronautics and Space Administration.




\end{document}